\documentclass[11pt]{amsart}
\usepackage{latexsym}
\usepackage{amssymb}
\usepackage{amsxtra}
\usepackage{amsmath}
\usepackage{amsfonts}

\newcommand{\cleqn}{\setcounter{equation}{0}}
\newcommand{\clth}{\setcounter{theorem}{0}} 
\newcommand {\sectionnew}[1]{\section{#1}\cleqn\clth}
\newcommand{\beq}{\begin{equation}}
\newcommand{\eeq}{\end{equation}}
\newcommand{\beqa}{\begin{eqnarray}}
\newcommand{\eeqa}{\end{eqnarray}}
\newcommand{\beaa}{\begin{eqnarray*}}
\newcommand{\ben}{\begin{eqnarray*}}
\newcommand{\eaa}{\end{eqnarray*}}
\newcommand{\een}{\end{eqnarray*}}

\newcommand \nc {\newcommand}

\newtheorem{theorem}{Theorem}[section]
\newtheorem{lemma}[theorem]{Lemma}

\newtheorem{corollary}[theorem]{Corollary}

\nc \thref{Theorem \ref}
\nc \leref{Lemma \ref}
\nc \prref{Proposition \ref}
\nc \coref{Corollary \ref}
\nc \deref{Definition \ref}
\nc \exref{Example \ref}
\nc \reref{Remark \ref}

\newcommand{\CC}{\mathcal{C}}

\newcommand{\W}{\mathcal{W}}
\newcommand{\A}{\mathcal{A}}
\newcommand{\B}{\mathcal{B}}
\newcommand{\C}{\mathbb{C}}
\newcommand{\D}{\mathcal{D}}

\newcommand{\F}{\mathcal{F}}
\renewcommand{\H}{\mathcal{H}}
\newcommand{\J}{\mathcal{J}}
\renewcommand{\L}{\mathcal{L}}
\newcommand{\M}{\mathcal{M}}

\newcommand{\QQ}{\mathbb{Q}}

\newcommand{\T}{\mathcal{T}}

\newcommand{\Z}{\mathbb{Z}}

\newcommand{\f}{\mathbf{f}}

\newcommand{\q}{\mathbf{q}}
\renewcommand{\t}{\mathbf{t}}
\newcommand{\x}{\mathbf{x}}
\newcommand{\y}{\mathbf{y}}

\def\Re{\mathop{\rm Re}\nolimits}

\def\dim{\mathop{\rm dim}\nolimits}

\def\d{\partial}
\def\ev{\mathop{\rm ev}\nolimits}

\def\iso{\cong}
\def\tensor{\otimes}

\def\Kahler{K\"ahler}
\def\Poincare{Poincar\'e}

\def\({\left(}
\def\){\right)}
\def\[{\left[}
\def\]{\right]}
\def\<{\left\langle}
\def\>{\right\rangle}

\def\lieGL{{\rm GL}}

\def\gl{\lambda}
\def\ge{\epsilon}
\def\ga{\alpha}
\def\gd{\delta}
\def\gb{\beta}
\begin{document}
\title{
Gromov--Witten theory of $\C P^1$ and integrable hierarchies}
\author{
Todor E. Milanov}
\thanks{E-mail: milanov@math.berkeley.edu}
\date{}

\begin{abstract}
The ancestor Gromov--Witten invariants of a compact {\Kahler}
manifold $X$ can be organized in a generating function called the 
total ancestor potential of $X$. 
In this paper, we construct Hirota Quadratic Equations (HQE shortly)
for the total ancestor potential of $\C P^1$. The idea is to adopt
the formalism developed in \cite{G1,GM} to the mirror model of $\C P^1$. 
We hope that the ideas presented here can be generalized to
other manifolds as well. 

As a corollary, using the twisted loop group formalism from \cite{G3}, 
we obtain a new proof of the following version of the Toda conjecture:
the total descendant potential of $\C P^1$ (known also as 
the partition function of the $\C P^1$ topological sigma model) 
is a tau-function of the Extended Toda Hierarchy.
\end{abstract}

\maketitle

\setcounter{section}{0}
\sectionnew{Introduction}
Let $X$ be a compact K\"ahler manifold. Denote $X_{g,m,d}$ the 
moduli space of degree $d\in H_2(X,\Z)$ {\em stable maps} to $X$
of genus-$g$ $m$-pointed complex curves. In case $X=pt$ the moduli
space is denoted by $\overline \M_{g,m}.$ 
The {\em total ancestor} potential of $X$ is defined as the following
generating function of Gromov--Witten invariants of $X$:
$$
\A_\tau = \exp\(\sum_{g\geq 0}\ge^{2g-2}\overline \F_\tau^{(g)}\), 
$$
where $\tau\in H^*(X;\C)$ is a parameter and $\overline \F_\tau^{(g)}$ are 
the genus-g ancestor potentials:
\ben
\overline \F_\tau^{(g)}=\sum_{l,m,d} \frac{Q^d}{l!m!}
\int_{[X_{g,l+m,d}]} 
\wedge_{i=1}^m \(\sum_{k\geq 0}\ev_i^*t_k \overline \psi_i^k\)
\wedge_{i=m+1}^{m+l} \ev_i^*\tau,
\een 
\indent $[X_{g,m,d}]$ --- the virtual fundamental class of $X_{g,m,d}$,\newline
\indent $t_n \in H:=H^*(X,\C [[Q]])$ --- 
arbitrary cohomology classes of $X$, \newline
\indent $\ev_j: X_{g,m,d} \to X$ --- evaluation at the $j$-th marked point,
\newline
\indent $Q^d$ --- the elements of the Novikov ring $\C [[Q]] := 
\C [[\ \text{Mori cone of $X$} ]]$, \newline
\indent
$\overline \psi_i = \pi^*(\psi_i),\ i=1,...,m,$ where the map 
$\pi:X_{g,m+l,d}\rightarrow\overline\M_{g,m+l}\rightarrow \overline \M_{g,m}$ 
is the composition of the contraction map with the operation of forgetting 
all markings except the first $m$ ones and $\psi_i$ are the 1-st
Chern classes of the universal cotangent lines over the moduli space 
of curves $\overline \M_{g,m}.$
\newline
It is a formal function in  the sequence of vector variables 
$t_0,t_1,t_2,...$ and $\tau.$ 

Let  $\H = H((z^{-1}))$ be the  space of formal Laurent series in $z^{-1}$,
equipped with the following symplectic structure:  
\beq\label{sympl_str}
\Omega(f,g):= \frac{1}{2\pi i} \oint
(f(-z),g(z))\ dz,
\eeq
where $(\ ,\ )$ is the {\Poincare} pairing on $H.$
The polarization $\H=\H_+\oplus \H_-$, defined by the Lagrangian subspaces
$\H_+ = H[z]$ and $\H_- =z^{-1} H[[z^{-1}]]$, identifies $\H$ with the 
cotangent bundle $T^*\H_+$.

Denote $\B$ {\em the Bosonick Fock space} which consists of functions on 
$\H_+$ which belong to the formal neighborhood
of $-{\bf 1}z$ (${\bf 1}$ is the unity in $H$) i.e., if we let 
$\q(z)=\sum_{k\geq 0} q_kz^k\in \H_+$ then $\B$ is the space of formal 
functions in the sequence of  vector variables 
$q_0,q_1+{\bf 1},q_2,\ldots .$ The total ancestor
potential $\A_\tau$ is identified with a vector in $\B$ via {\em the
dilaton shift} $\t(z)=\q(z)+z,$ where $\t(z)=\sum t_kz^k$. 

Any $\f\in\H$ can be written uniquely as 
\ben
\f=\sum_{k\geq 0,a} 
q_{k,\ga}\ \phi_a z^k+p_{k,a}\ \phi^{a}(-z)^{-1-k}, 
\een 
where $\{\phi_a\}$ is a basis of $H$ and $\{\phi^a \}$ is its dual 
with respect to the {\Poincare} pairing. The coefficients $p_{k,a}$
$q_{k,a}$ are coordinate functions on $\H$ which form a Darboux 
coordinate system. Thus the formulas 
\beq
\label{qlinear}
\hat q_{k,a} := {q_{k,a}}/{\ge},\quad
\hat p_{k,a} := \ge {\d}/{\d q_{k,a}}
\eeq
define a representation of the Heisenberg 
Lie algebra generated by the linear Hamiltonians on the Fock space $\B.$
Given a vector ${\bf f}\in \H$ we define 
{\em a vertex operator} acting on $\B$: 
$e^{\widehat \f}:=(e^\f)\sphat:=e^{\hat \f_-}e^{\hat \f_+} ,$ where 
$\f_\pm$ is the projection of $\f$ on $\H_\pm$ and $\f_\pm$ is 
identified with the linear Hamiltonian $\Omega(\cdot ,\f_\pm ).$ 

The far reaching goal is to construct HQE for the total ancestor 
potential of $X$ in terms of vertex operators acting on the Fock
space $\B.$ In this paper we will solve this problem for 
$X=\C P^1.$ The vertex operators will be defined through the mirror 
model of $\C P^1$ and it is not hard to generalize the definition
to other manifolds as well. However, it is hard to generalize the HQE.  

When $X=$pt, the total ancestor potential is in fact independent
of the parameter $\tau\in H^*({\rm pt},\C)\iso \C$ (see subsection
3.1) and it will be
denoted by $\A_{\rm pt}.$ By definition, $\A_{\rm pt}$ is the total 
ancestor potential of $A_1$ singularity. 
Thus, according to the Corollary of Proposition 2 in \cite{G1}, $\A_{\rm pt}$ 
satisfies a family of HQE parametrized by $\tau\in \C.$ Let us write 
down this family of HQE explicitly. We will take a slightly different 
point of view from \cite{G1} which will be used also for the case 
$X=\C P^1.$ 

Let 
$f_\tau:\C \rightarrow \C$ be the function $f_\tau(x)=x^2/2+\tau.$ For
fixed $\tau$, $f_\tau$ is a Morse function with critical point 
$x_{1}=0$ and critical value $u_1=\tau$. Pick a reference point  
$\gl_0\in \C\backslash \{u_1\}$ and a path
$C_1$ connecting $\gl_0$ with $u_1$ and such that it approaches $u_1$
along a straight segment. Denote by $\gb$ the corresponding 
{\em Lefschetz thimble} i.e., the relative 
homology cycle in $H_1(\C, f_\tau^{-1}(\gl_0);\Z)$ represented by 
$f_\tau^{-1}(C_1)$ with an arbitrary choice of the orientation. For
each $n\in \Z$ we define a period vector $I_\gb^{(n)}:$ 
\ben
I^{(-k)}_\gb(\gl,\tau) & = &
(\d/\d \gl)\int_{\gb(\gl)} \frac{(\gl-f_\tau(x))^{k}}{k!}\omega  ,  \\
I^{k}_\gb(\gl,\tau)  & = & \d_\gl^{k}I_\gb^{(0)}(\gl,\tau),
\een
where $k$ is a non-negative integer, $\omega =dx,$ and 
$\gb(\gl)\in H_1(\C, f_\tau^{-1}(\gl);\Z)$ is a Lefschetz thimble 
obtained by extending $\gb$ along a path $C$ connecting $\gl_0$ 
and $\gl.$ The period vectors are multivalued 
functions on  $\C\backslash \{u_1\}$ with values in $H.$

If $\ga=r\gb$ for some $r\in \QQ,$ then we define: 
$$
I_\ga^{(n)}=rI_\gb^{(n)},\ \ \ 
\f_\tau^\ga(\gl)=\sum_n I_\ga^{(n)}(\gl,\tau)(-z)^n, \ \ \ 
\Gamma_\tau^\ga =e^{\widehat\f_\tau^\ga}.$$
Let $\ga=\gb/2.$ Then the integrals defining the corresponding 
period vectors can be computed explicitly and we get:
\beq\label{vop_kdv}
\Gamma^{\ga}_\tau = \exp \left(
\pm\sum_{n\geq 0} \frac{(2(\gl-\tau))^{n+1/2}}{(2n+1)!!}\frac{q_n}{\ge}
\right)
\exp\left(
\mp\sum_{n\geq 0}\frac{(2n-1)!!}{(2(\gl-\tau))^{n+1/2}}\ge \d_{q_n} \right),
\eeq
where the sign depends on the choice of the path $C.$ 
We remark that 
\beq \notag 
\frac{(2n-1)!!}{(2(\gl-\tau))^{1/2+n}} = \left(-\frac{d}{d\gl}\right)^n 
\frac{1}{\sqrt{2(\gl-\tau)}}, \ \ \frac{(2(\gl-\tau))^{1/2+n}}{(2n+1)!!} = 
\left(\frac{d}{d\gl}\right)^{-1-n} \frac{1}{\sqrt{2(\gl-\tau)}} .\eeq
Let
\ben
c_\ga(\gl,\tau) := \lim_{\ge \rightarrow 0}\ 
\exp\( \int^{u_1+\ge}_\gl 
\(I^{(0)}_\ga(\xi,\tau),I^{(0)}_\ga(\xi,\tau)\)d\xi
-\langle \ga,\ga \rangle\,\int_1^\ge\frac{d\xi}{\xi}  \) = 
\frac{1}{\sqrt{\gl-\tau}}, 
\een
where $\langle\ ,\ \rangle$ is the intersection pairing in 
$H_1(\C,f_\tau^{-1}(\gl);\Z):$  $\langle \gb,\gb \rangle =2.$
Here, the limit $\ge\rightarrow 0$ is taken along a straight segment
such that $u_1+\ge$ parametrizes the end of the path $C_1.$ 
The integration path in the 1-st integral is $C_1(\ge)\circ C^{-1},$
where $C_1(\ge)$ is the path obtained from $C_1$ by truncating the
line segment between $u_1+\ge$ and $u_1,$ and in the 2-nd one --- 
a straight segment between 1 and $\ge.$

The HQE for the ancestor potential of a point can be stated this way:
\beq\label{KdV}
c_{\ga}\( 
\Gamma_\tau^{\ga}\tensor \Gamma_\tau^{-\ga} - 
\Gamma_\tau^{-\ga}\tensor \Gamma_\tau^{\ga}\) 
\(\A_{\rm pt}\tensor\A_{\rm pt}\)d\gl \ \ \mbox{ is regular in }\ \ \gl. 
\eeq
Here $\A_{\rm pt} \otimes A_{\rm pt}$  means the function 
$A_{\rm pt} (\q') A_{\rm pt} (\q'')$
of the two copies of the variable $\q = (q_0,q_1,...)$, and the vertex
operators in $\Gamma_\tau^{\pm\ga}\otimes\Gamma_\tau^{\mp\ga}$ preceding 
(respectively --- following) $\otimes$ act on $\q'$ 
(respectively --- on $\q''$). 
The expression in (\ref{KdV}) is in fact single-valued for $\gl$ near
$\infty.$ 
Passing to the variables $\x = (\q'+\q'')/2$ and 
$\y = (\q'-\q'')/2$ and using Taylor's formula one can expand (\ref{KdV}) 
into a power series in $\y$ with coefficients which are Laurent series 
in $\gl^{-1}$ (whose coefficients are polynomials in $\A_{\rm pt}$ and its
partial derivatives). The regularity condition in (\ref{KdV}) means,
by definition,  that all the Laurent series in $\gl^{-1}$ are polynomials
in $\gl$.

The HQE \eqref{KdV} for $\A_{\rm pt}$ are consequence of Witten's conjecture
\cite{W}, proved by Kontsevich \cite{Ko}, and the string equation.

\medskip

Let $X=\C P^1$ and $\phi_0={\bf 1}$, $\phi_1= P$ ($P$ is the 
cohomology class of the hyperplane section of $\C P^1$) be a 
basis in $H=H^*(\C P^1;\C)$. We will assume that the parameter
of the ancestor potential is $\tau = t P.$ Let 
$f_t:\C^*\rightarrow \C$ be the function  $f_t(x)=x+(Qe^{t}/{x}).$
For fixed $t$, $f_t$ is a Morse function with critical points 
$x_{1/2}=\pm\sqrt{Q}e^{t/2}$ and critical values 
$u_{1/2}=\pm2\sqrt{Q}e^{t/2}.$ Pick a reference point  
$\gl_0\in \C\backslash \{u_1,u_2\}$ and paths
$C_i,\ i=1,2$ connecting $\gl_0$ with $u_i$ and such that $C_i$
approaches $u_i$ along a straight segments. Denote 
$\gb_i$ the corresponding {\em Lefschetz thimbles} i.e., the relative 
homology cycles in $H_1(\C^*, f_t^{-1}(\gl_0);\Z)$ represented by 
$f_t^{-1}(C_i)$. We fix the orientation on $\gb_i$ in such a way that
$\phi:=\gb_1+\gb_2$ is a cycle homologous to the circle (with the counter-
clockwise orientation) around the puncture of the punctured plane $\C^*.$ 
For each $n\in \Z$ we define period vectors $I_{\gb_i}^{(n)}:$ 
{\allowdisplaybreaks
\ben
(I^{(-k)}_{\gb_i}(\gl,t),1) & = &
(\d/\d \gl)\int_{\gb_i(\gl)} \frac{(\gl-f_t(x))^{k}}{k!}\omega  ,  \\
(I^{(-k)}_{\gb_i}(\gl,t),P) & = &
-(\d/\d t)\int_{\gb_i(\gl)} \frac{(\gl-f_t(x))^{k}}{k!}\omega  ,  \\
I^{k}_{\gb_i}(\gl,t)  & = & \d_\gl^{k}I_{\gb_i}^{(0)}(\gl,t),
\een }
where $k$ is a non-negative integer, $\omega =dx/x,$ and 
$\gb_i(\gl)\in H_1(\C^*, f_t^{-1}(\gl);\Z)$ is a Lefschetz thimble 
obtained by extending $\gb_i$ along a path $C$ connecting $\gl_0$ 
and $\gl.$ The period vectors are multivalued 
functions on  $\C\backslash \{u_1,u_2\}$ with values in $H.$

If $\ga=r_1\gb_1+r_2\gb_2$ for some $r_1,r_2\in \QQ,$ then we
define: 
$$
I_\ga^{(n)}=r_1I_{\gb_1}^{(n)}+r_2I_{\gb_2}^{(n)},\ \ \ 
\f_\tau^\ga(\gl)=\sum_n I_\ga^{(n)}(\gl,t)(-z)^n, \ \ \ 
\Gamma_\tau^\ga =e^{\widehat\f_\tau^\ga}.$$
Let $\ga=\gb_1/2.$ Introduce the function
\ben
c_\ga(\gl,\tau) = \lim_{\ge \rightarrow 0}\ 
\exp\( \int^{u_1+\ge}_\gl 
\(I^{(0)}_\ga(\xi,\tau),I^{(0)}_\ga(\xi,\tau)\)d\xi
-\langle \ga,\ga \rangle\,\int_1^\ge\frac{d\xi}{\xi}  \), 
\een
where $\langle\ ,\ \rangle$ is the intersection pairing in 
$H_1(\C^*,f_\tau^{-1}(\gl);\Z):$ 
$\langle \gb_1,\gb_1 \rangle =\langle \gb_2,\gb_2 \rangle= 2$ and 
$\langle \gb_1,\gb_2 \rangle=-2$. The limit $\ge\rightarrow 0$ is   
taken along a straight segment such that $u_1+\ge$ 
parametrizes the end of the path $C_1.$ 
The 1-st integration path is the path $C_1(\ge)\circ C^{-1},$ where
$C_1(\ge)$ is obtained from $C_1$ by truncating the line segment
between $u_1+\ge$ and $u_1,$ and the 2-nd one --- a straight segment
between 1 and $\ge.$ 

The expression similar to \eqref{KdV} is not single valued 
near $\infty$ -- we will prove that 
the vertex operators $\Gamma_\tau^{\pm\ga}\tensor\Gamma_\tau^{\mp\ga}$
under the analytic continuation $\gl=\infty$ are multiplied by
the monodromy factors $\Gamma_\tau^{\pm\phi}\tensor\Gamma_\tau^{\mp\phi}.$
To offset the complication we will generalize the concept  
of vertex operators.

We will allow vertex operators with coefficients in the algebra 
$\A$ of differential operators 
$\sum_{0\leq k \leq N} a_k(x;\ge)\d_x^k,$ where each
$a_k$ is a formal Laurent series in $\ge$ with coefficients smooth
functions in $x.$ We equip $\A$ with the {\em anti-involution}
$\#$ which acts on the generators of $\A$ as follows:
$$
(\ge\d_x)^\# = -\ge\d_x +\log Q,\ \ \ x^\# = x.
$$
Let $w_\tau$ and $v_\tau$ be the vectors in 
$\H$ defined by:
\beqa\label{wt}
&& w_\tau   = -Pz^{-1} - 
\sum_{k\geq 0\, , a=0,1} 
\langle P\psi^k,\phi_a\rangle_{0,2}(\tau)\phi^a z^{-k-2}, \\
\label{vt}
&& v_\tau = 
\ \ \ {\bf 1}      \ \ \ \ \ \ +
\sum_{k\geq 0\, , a=0,1} 
\langle \psi^k,\phi_a\rangle_{0,2}(\tau)\phi^az^{-k-1} ,
\eeqa 
where the correlator notation stands for
\ben
\langle \phi_\ga\psi^k,\phi_\gb\rangle _{0,2}(\tau)=
\sum_{n,d}\frac{Q^d}{n!}\int_{[X_{0,n+2,d}]}
\ev_1^*(\phi_\ga)\psi_1^k\wedge\ev_2^*(\phi_\gb)\wedge
(\wedge_{i=3}^{n+2}\ev_i^*(\tau))\ ,
\een
where $\psi_1$ is the 1-st Chern class of the cotangent line (at the
1-st marked point) bundle over the moduli space $X_{0,n+2,d}.$

Introduce the vertex operator, acting on the algebra  
$$
\pi:=\B\tensor_{\C((\ge))}\A=\left\{ 
\sum_{k=1}^N a_k(x,\q;\ge)\d_x^k\  \right\}$$
of differential operators with coefficients in the Fock space $\B,$   
\ben
\Gamma_\tau^\delta = 
\exp\left( 
(\hat \f_\tau^{\phi/2\pi i}-\hat w_\tau)(\ge\d_x-\log \sqrt Q) \right)
\exp ( (x/\ge)\hat v_\tau) .
\een 
We define the following HQE: {\em a function $\T\in \B$ satisfies the
HQE of $\C P^1$ if}
\beqa
\label{HQE_a}
&&
\(\Gamma_\tau^{\gd\#}\tensor \Gamma_\tau^\gd \) 
c_\ga \( \Gamma_\tau^{\ga}\tensor \Gamma_\tau^{-\ga} -
  \Gamma_\tau^{-\ga}\tensor \Gamma_\tau^{\ga}\)
\(\T\tensor \T\)\  {d\gl}
\eeqa
{\em computed at $\q'$ and $\q''$ such that
$\hat w_\tau' - \hat w_\tau'' =m$, is regular in 
$\gl$ for each $m\in \Z$.}

The expression \eqref{HQE_a} is interpreted as taking values in the algebra
$\pi\tensor_\A\pi$ of differential operators with coefficients depending on 
$\q',\q'',\ge$ and $\gl$.
We will show that for every $r\in \QQ$ the following identity
holds:
\ben
\(\Gamma_\tau^{\delta \#}\tensor\Gamma_\tau^\delta\)
\(\Gamma_\tau^{r\,\phi}\tensor  \Gamma_\tau^{-r\,\phi}\) =
e^{2\pi i(\hat w_\tau\tensor 1 - 1\tensor \hat w_\tau)\,r}
\Gamma_\tau^{\delta \#}\tensor\Gamma_\tau^\delta.
\een
Thus when 
$\hat w_\tau'-\hat w_\tau'' \in \Z,$ the expression
\eqref{HQE_a} is single-valued near $\gl=\infty.$ After the change 
$\y=(\q'-\q'')/2, \x=(\q'+\q'')/2$ and the substitution 
\footnote{according to our quantization rules: 
$\hat w_\tau = -q_{0,0}/\ge+
\sum_{k,a}(-1)^k\<P\psi^k,\phi_a\>_{0,2}(\tau)q_{k+1,a}/\ge$} 
$$
y_{0,0} = -\frac{m\ge}{2} + \sum_{k= 0}^\infty\sum_{a=0,1} 
\langle P\psi^k,\phi_a\rangle_{0,2}(\tau)(-1)^{k+1}y_{k+1,a}
$$
it expands (for each integer $m$) 
as a power series in $\y$ ($y_{0,0}$ excluded) with coefficients 
which are Laurent series in $\gl^{-1}$ 
(whose coefficients are {\em differential operators in $x$}
depending on $\x$ via $\T$, its translations and partial derivatives). 
\begin{theorem}\label{t1}
The total ancestor potential of $\C P^1$ satisfies the HQE \eqref{HQE_a}.
\end{theorem}
 
\medskip

As an application of \thref{t1}, we will give a proof of the so called
Toda conjecture (see \cite{EY,Ge1,OP2,CDZ, Z}) about Gromov -- Witten 
invariants of $\C P^1$. 

Let $X$ be a compact {\Kahler} manifold. The {\em total descendant potential}
of $X$ is defined as the following generating function for Gromov -- Witten 
invariants of $X$:
\ben 
\D_X :=\exp \sum_{g\geq 0} \ge^{2g-2} \F^{(g)}, \ \ 
\F^{(g)} := \sum_{m,d}\frac{Q^d}{m!} \int_{[X_{g,m,d}]} \prod_{j=1}^m
\sum_{n\geq 0} \ev_j^*(t_n) \psi_j^n, \een 
where $\psi_j$ are the 1-st Chern classes of the universal 
cotangent lines over $X_{g,m,d}$ and $t_n \in H^*(X,\QQ [[Q]])$ are 
arbitrary cohomology classes of $X$. 
It is identified with a vector in the Fock space $\B$ via the 
dilaton shift $\t(z)=\q(z)+z$. 
\begin{corollary}\label{c}
The total descendant potential of $\C P^1$ is a tau-function of the
Extended Toda Hierarchy.
\end{corollary}  
The Toda Conjecture was suggested by T. Eguchi and S.-K. Yang \cite{EY}. 
The original formulation was inaccurate. The correct one was given by
E. Getzler \cite{Ge1} and independently by Y. Zhang \cite{Z}.

Apparently, the first proof was obtained 
by E. Getzler \cite{Ge1} who applied two results -- the unextended part
of the Toda conjecture and the Virasoro constraints for $\C P^1$. 
The paper \cite{DZ2} by B. Dubrovin -- Y. Zhang contains a different 
proof based on the theory \cite{DZ1} of integrable hierarchies 
associated to Frobenius structures and also uses the Virasoro constraints 
for $\C P^1$. The Virasoro constraints for
$\C P^1$ were proved by A. Givental in \cite{G3} by combining the
fixed point localization formula \cite{G4} for $\D_X$ with mirror symmetry
and a certain loop group formalism. Our proof to the Toda Conjecture 
stays entirely within the paradigm developed in \cite{G3} and  
pursued further in \cite{G1, GM}. It relies directly on this formalism 
and the mirror model of $\C P^1$ as well as Kontsevich's theorem. 
Yet another approach to the Toda conjecture, due to A. Okounkov -- 
R. Pandharipande \cite{OP2}( and based on fixed point localization but 
buy-passing Kontsevich's theorem), yields the {\em equivariant} version 
of the Toda conjecture \cite{Ge2} as well as the unextended part
of the non-equivariant Toda conjecture.

\medskip

{\bf Acknowledgments.} I am thankful to A. Givental for teaching me the
methods from \cite{G1} and for the many stimulating discussions.

\sectionnew{From Ancestors to KdV}
 
In this section we give a proof of \thref{t1}. 
\subsection{The HQE of $\C P^1$} The vertex oerators 
$\Gamma_\tau^{\pm\ga}(\gl)$ and $\Gamma^\delta_\tau,$ and 
the function $c_\ga(\gl,\tau)$ depend on a choice of a path $C$ 
connecting $\gl$ with the reference point $\gl_0.$ We want
to show that the expression \eqref{HQE_a}, when computed
at $\widehat w_\tau'-\widehat w_\tau'' \in \Z,$ is independent
of the choice of $C.$
   
Recall that the thimbles $\gb_i,\ i=1,2$ are determined by the
paths $C_i,\ i=1,2.$ We fix generators $\gamma_i$ of the 
fundamental group $\pi_1(\C\backslash \{u_1,u_2\},\gl_0)$ 
as follows: $\gamma_i$ is a path starting at $\gl_0$ and approaching
$u_i$ along the path $C_i$; when sufficiently close to $u_i$, $\gamma_i$
makes a small circle (counter-clockwise) around $u_i$ and  
it returns back to $\gl_0$ along the path $C_i.$ 
Denote $C'=C\circ\gamma_i$ and by $\gb_1'(\gl)$ and $\gb_2'(\gl)$ 
the Lefschetz thimbles corresponding to the new path $C'.$ 
For simplicity, we will consider the case $i=2$ (the case
$i=1$ is similar and simpler).
 
According to the Pickard--Lefschetz theorem, the change of
the thimbles is measured by the following formula (\cite{AGV},
chapter 1, section 1.3):  
\ben
&&
\gb_1'(\gl) = \gb_1(\gl) - \langle \gb_1 ,\gb_2 \rangle \gb_2(\gl) = 
\gb_1(\gl)+2\gb_2(\gl), \\
&&
\gb_2'(\gl)  = \gb_2(\gl) - \langle \gb_2 ,\gb_2 \rangle \gb_2(\gl) = 
-\gb_2(\gl). 
\een
Thus 
\ben
\ga'(\gl)  & = & \gb_1'/2 = -\ga(\gl) + \phi(\gl), \\
\phi'(\gl) & = & \gb_1'(\gl)+\gb_2'(\gl) =  \gb_1(\gl)+\gb_2(\gl) = \phi(\gl)
\een
and we get
$\f_\tau^{\pm\ga'}(\gl) =\f_\tau^{\mp\ga}(\gl) + \f_\tau^\phi(\gl)$
and $ \f_\tau^{\phi'}(\gl)=\f_\tau^\phi(\gl).$ In particular, we find
that the vertex operator $\Gamma_\tau^\delta$ does not depend on the
choice of $C.$ Let us compare $\Gamma_\tau^{\pm \ga'}$ and 
$\Gamma_\tau^{\pm \ga}.$ Note that
\beqa\label{I_phi}
(I_\phi^{(-1)}(\gl,\tau),1) &=&
\d_\gl \int_\phi \(\gl-f_t(x)\)dx/x = 2\pi i, \\ \label{I_phii}
(I_\phi^{(-1)}(\gl,\tau),P) &=&
-\d_t \int_\phi \(\gl-f_t(x)\)dx/x = 0. 
\eeqa
Thus $I_\phi^{(k)} = 0$ for all $k\geq 0$ i.e., $(\f^\phi_\tau)_+=0$
and we get 
$\Gamma_\tau^{\pm \ga'} = \Gamma_\tau^{\pm\phi}\Gamma_\tau^{\mp\ga}.$ 

Let us analyze the coefficients $c_\ga.$ 
Introduce the following vectors in $H$\footnote{
$\{{\bf 1}_1,{\bf 1}_2\}$ is a basis of $H$ such that 
$({\bf 1}_i,{\bf 1}_j)=\delta_{ij}$ and 
${\bf 1}_i\bullet_\tau{\bf 1}_j=\delta_{ij}{\bf 1}_i,$ where $\bullet_\tau$
is the quantum cup product at the point $\tau = tP\in H.$}:
$$
{\bf 1}_i = \frac{1}{\sqrt\Delta_i}\(\phi^0 + \frac{\d u_i}{\d t}\phi^1\), 
\ i=1,2,$$ 
where, recall that $\{\phi_0={\bf 1},\phi_1=P\}$ is a basis of $H,$ 
$\{\phi^0,\phi^1\}$ is its dual  basis 
with respect to the {\Poincare} pairing, and $u_i(t)$ are the critical 
values of the function $f_t(x)=x+Qe^t/x.$  The factors $\Delta_i$ are
the Hessians of $f_t$ at the critical point $x_i$ with respect to the 
volume form $\omega.$

\begin{lemma}\label{I_ui}
For $\xi$ sufficiently close to $u_i$ the period vector $I^{(0)}_{\gb_i}$
can be expanded into a series of the following type
\beq\label{i_0}
I^{(0)}_{\gb_i}(\xi,\tau) = 
\frac{2}{\sqrt{2(\xi-u_i)}}\({\bf 1}_i + A_{i,1}[2(\xi-u_i)]+ 
A_{i,2}[2(\xi-u_i)]^2+\ldots\ \), 
\eeq
where the path $C$ specifying $\gb_i(\xi)$ is the same as $C_i$ everywhere
in $\C\backslash\{u_1,u_2\},$ except for a small disk around $u_i,$
where the two paths split -- $C_i$ leads to $u_i$ and $C$ leads to $\xi.$  
\end{lemma} 
\proof
We follow \cite{AGV}, chapter 3, section 12, Lemma 2. 

Let $y$ be {\em a unimodular} coordinate on $\C^*$ for the volume form 
$\omega$ i.e., $\omega = dy.$ The Taylor's expansion of $f_t$ is:
\ben
f_t(y) = u_i + \frac{\Delta_i}{2}(y-y_i)^2 + \ldots ,
\een
where $y_i$ is the $y$-coordinate of the critical point $x_i.$ From this
expansion we find that the equation $f_t(y) = \xi$ has two 
solutions in a neighborhood of $\xi=u_i$:
\ben
y_\pm = y_i \pm \frac{1}{\sqrt\Delta_i} \sqrt{2(\xi-u_i)} + \mbox{ h.o.t.},
\een
where h.o.t. means higher order terms. Thus the integral in the definition
of $I_{\gb_i}^{(0)}(\xi,\tau)$ has the following expansion:
\ben
\int_{\gb_i(\xi)}\omega = y_+(\xi) - y_-(\xi) = 
\frac{2}{\sqrt\Delta_i} \sqrt{2(\xi-u_i)} + \mbox{ h.o.t. }.
\een
The last expansion yields:
\ben
&&
(I_{\gb_i}^{(0)}(\xi,\tau),1) = 
\frac{2}{\sqrt{2(\xi-u_i)}}\, 
\frac{1}{\sqrt\Delta_i} + \mbox{ h.o.t. },\\
&&
(I_{\gb_i}^{(0)}(\xi,\tau),P) = 
\frac{2}{\sqrt{2(\xi-u_i)}}\, 
\frac{1}{\sqrt\Delta_i}\frac{\d u_i}{\d t} + \mbox{ h.o.t. }.
\een
The lemma follows.
\qed

Using \leref{I_ui} we get 
\beq\label{change_ca}
c_{\ga'}/c_{\ga} = \exp\(-\int_{\gamma_2} 
(I^{(0)}_\ga(\xi,\tau),I^{(0)}_\ga(\xi,\tau))d\xi\) = 
\exp\(-\frac{1}{2}\int_{\gamma_2}\frac{d\xi}{\xi-u_i} \) = -1.
\eeq 
For the 2-nd equality in \eqref{change_ca}, we used that 
$I^{(0)}_{\ga} = I^{(0)}_{\phi/2} - I^{(0)}_{\gb_2/2} = 
- I^{(0)}_{\gb_2/2}$ and
that the expansion \eqref{i_0} holds (because $\gamma_2$ is a small
loop around $u_2$) and only its  leading term contributes to the integral.   

\begin{lemma}{\label{OPE1}}
Let $r\in \QQ.$
Then the following identity between operators acting on $\pi\tensor_\A\pi$
holds:
\ben
\(\Gamma_\tau^{\delta \#}\tensor\Gamma_\tau^\delta\)
\(\Gamma_\tau^{r\,\phi}\tensor  \Gamma_\tau^{-r\,\phi}\) =
e^{2\pi i(\hat w_\tau\tensor 1 - 1\tensor \hat w_\tau)\,r}
\(\Gamma_\tau^{\delta \#}\tensor\Gamma_\tau^\delta\).
\een
\end{lemma}
\proof
By definition (see formulas \eqref{I_phi})  
\ben
\f_\tau^\phi(\gl) = 2\pi i\, P (-z)^{-1} + 
\sum_{k\geq 1} I_\phi^{(-1-k)}(\gl,\tau)(-z)^{-k-1}.
\een
Comparing with the definition of $w_\tau$ we get
$\f_\tau^{\phi/2\pi i}(\gl)\in w_\tau+z^{-1}\H_-.$ Thus
\ben
\Omega(v_\tau,\f_\tau^{\phi/2\pi i}(\gl)) = \Omega(v_\tau,w_\tau)=-1,\ \ 
\Omega(w_\tau,\f_\tau^{\phi/2\pi i}(\gl)) =0 \ .
\een

For all $f,g\in \H$ we have: 
\ben
e^{\hat f}e^{\hat g} = e^{[\hat f,\hat g]}e^{\hat g}e^{\hat f} = 
                       e^{\Omega(\hat f,\hat g)}e^{\hat g}e^{\hat f},
\een
because for linear Hamiltonians the quantization is a representation of
Lie algebras. 
Thus
{\allowdisplaybreaks
\ben
&&
\Gamma_\tau^\delta\Gamma_\tau^{r\,\phi} = 
\exp\left( 
(\widehat \f_\tau^{\phi/2\pi i}-\widehat w_\tau)(\ge\d_x-\log \sqrt Q) \right)
\exp \( \frac{x}{\ge}\widehat v_\tau\)\exp\(r\widehat\f_\tau^\phi\)= \\
&&
\exp\(r\widehat\f_\tau^\phi\)\exp\left( 
(\widehat \f_\tau^{\phi/2\pi i}-\widehat w_\tau)(\ge\d_x-\log \sqrt Q) \right)
\exp\((r x/\ge)\Omega(v_\tau,\f_\tau^\phi)\)\times \\
&&
\exp \( \frac{x}{\ge}\widehat v_\tau\)=
e^{2\pi i(\hat w_\tau - (x/\ge))r} \Gamma_\tau^\delta
\een}
Similarly,
$
\Gamma_\tau^{\delta\#}\Gamma_\tau^{r\,\phi} = 
e^{2\pi i\hat w_\tau\,r} \Gamma_\tau^{\delta\#} e^{-2\pi i x r/\ge}.
$
The Lemma follows. 
\qed

\medskip
Since $\Gamma_\tau^{\pm\ga'} = \Gamma_\tau^{\pm\phi}\Gamma_\tau^{\mp\ga}$ and
$c_{\ga'}=-c_\ga$, using \leref{OPE1}, we get that the HQE of $\C P^1$ 
do not depend on the choice of the path $C.$

\subsection{Tame asymptotical functions}
The total ancestor potential has some special property
which makes the expression \eqref{HQE_a} a formal series 
with coefficients meromorphic functions in $\gl.$

An {\em asymptotical function} is, by definition, an expression 
\ben
\T = \exp \sum_{g=0}^\infty \ge^{2g-2}\T^{(g)}(\t;Q),
\een
where $\T^{(g)}$ are formal series in the sequence of vector 
variables $t_0,t_1,t_2,\ldots$  with coefficients in the Novikov
ring $\C[[Q]].$ Furthermore, $\T$ is called {\em tame} if
\ben
\left.\frac{\d}{\d t_{k_1,a_1}} \ldots \frac{\d}{\d t_{k_r,a_r}}\right|_{\t=0} 
\T^{(g)} = 0 \quad \mbox{whenever} \quad
k_1+k_2+\ldots +k_r > 3g-3+r,
\een
where $t_{k,a}$ are the coordinates of $t_k$ with respect to 
$\{\phi_0,\phi_1\}.$
The total ancestor potential $\A_\tau$ is a tame asymptotical 
function, because the tameness conditions is trivially satisfied
for dimensional reasons: $\dim \overline \M_{g,r}=3g-3+r.$ 

Let $\T$ be a tame asymptotical function.  
The dilaton shift $\t(z)=\q(z)+z$ identifies $\T$ with an element
of the Fock space. Let 
$\Gamma_i^{\pm} = \exp\(\pm\sum I_i^{(n)}(\gl)(-z)^n\)$ be a finite
set of vertex operators, where $I_i^{(n)}$ are meromorphic functions. 
Consider the expression 
\beq\label{tameness}
\sum_i c_i(\gl)\(\Gamma_i^+\tensor\Gamma_i^-\)\(\T(\q')\T(\q'')\)d\gl, 
\eeq
where $c_i$ are meromorphic functions. 
According to \cite{G1}, Proposition 6, the tameness of $\T$ implies that
\eqref{tameness}, after the substitution $\q'=\x+\ge \y$, 
$\q''=\x-\ge\y$ and after dividing by $\exp\(2\T^{(0)}(\x)/\ge^2\),$
expands into a power series in $\ge,$ $\x,$ and $\y$ whose 
coefficients depend polynomially on finitely many $I_i^{(n)}.$

In particular, \eqref{HQE_a} can be interpreted as a formal series 
in $\ge$, $\x,$ and $\y$ ($y_{0,0}$ excluded) with coefficients 
meromorphic functions in $\gl.$ The vertex operators could have 
poles only at the critical values $u_1,u_2.$ Thus the 
regularity property follows if we prove that there are no poles at
$\gl=u_i.$   

\subsection{The twisted loop group formalism}
Let 
$$
\L^{(2)}\lieGL(H) = \left\{
M(z)\in \lieGL(\H)\ |\ M^*(-z)M(z)=1\right\},
$$
where $*$ means the transposition with respect to the {\Poincare} pairing, 
be the twisted loop group. 
The elements of the twisted loop group of the type
$M=1+M_1z+M_2z^2+\ldots$ (respectively $M=1+M_1z^{-1}+M_2z^{-2}+\ldots)$
are called upper-triangular (respectively lower-triangular) linear 
transformations. 
They can be quantized as follows: write $M=\log A,$ then $A(z)$ is an
infinitesimal symplectic transformation. We define  
$\widehat M=\exp \hat A,$ where $A$ is identified with the 
quadratic Hamiltonian $\Omega(A\f,\f)/2$ and on the space of
quadratic Hamiltonians the quantization rule $\sphat\ $ is defined by:
\ben
(q_{k,\ga}q_{l,\gb})\sphat := \frac{q_{k,\ga}q_{l,\gb}}{\ge^2},\ \ 
(q_{k,\ga}p_{l,\gb})\sphat := q_{k,\ga}\frac{\d}{\d q_{l,\gb}},\ \ 
(p_{k,\ga}p_{l,\gb})\sphat := \ge^2\frac{\d^2}{\d q_{k,\ga}\d q_{l,\gb} }.
\een 
We remark that $\sphat\ $ defines only {\em a projective representation} of
the subgroups of lower-triangular and upper-triangular elements of 
$\L^{(2)}\lieGL(H)$ on the Fock space $\B.$

For the proof of \thref{t1} we will need to conjugate vertex operators
with an upper-triangular linear transformation 
$R=1+R_1z+R_2z^2+\ldots\in\L^{(2)}\lieGL(H).$ The following
formula holds (\cite{G1}, section 7):
\beq\label{conj_R}
\widehat R^{-1} e^{\hat f} \widehat R = 
e^{V f_-^2/2}\(e^{R^{-1}f}\)\sphat,
\eeq 
where $-$ means truncating the non-negative powers of $z,$ $f_-=
\sum_{k\geq 0} (-1)^{-1-k}(f_{-1-k},q_k)$ is
interpreted (via the symplectic form) as a linear function in $\q,$ and 
$V(\d,\d) = \sum (V_{kl}\phi^a,\phi^b)\d_{q_{l,a}}\d_{q_{k,b}}$ is 
a second order differential operator whose coefficients are defined by
\beq\label{conj_v}
\sum_{k,l\geq 0} V_{kl}w^kz^l = \frac{1-R(w)R^*(z)}{w+z},
\eeq

\subsection{The ancestor potential and mirror symmetry}
Following \cite{G3} we will use the twisted loop group formalism to
express the total ancestor potential in terms of oscillating
integrals defined on the mirror model of $\C P^1.$

Let $\{e_1,e_2\}$ be the standard basis of $\C^2.$ Equip the loop 
space $\C^2((z^{-1}))$ with a symplectic structure via \eqref{sympl_str},
where the inner product in $\C^2$ is the standard one:
$(e_i,e_j)=\delta_{ij}.$ Denote $\B_{\C^2}$ the Bosonic Fock space 
which consists of functions defined in the formal neighborhood of 
$-e_1-e_2.$ In fact, $\C^2((z^{-1}))\iso
\H_{\rm pt}\oplus\H_{\rm pt} $ and  $\B_{\C^2}\iso
\B_{\rm pt}\tensor\B_{\rm pt},$ where $\H_{\rm pt} $  and $\B_{\rm pt}$  
are respectively the symplectic vector space and the Bosonic Fock
space associated with $X={\rm pt}.$

On the other hand, the map $\Psi(t):\C^2\rightarrow H,$
defined by $\Psi(e_i):={\bf 1}_i,$ is a linear isomorphism
which respects the inner products in $\C^2$ and $H.$ Thus
$\Psi(t)$ induces isomorphisms 
$\Psi(t):\C^2((z^{-1}))\rightarrow \H$ and 
$\widehat \Psi: \B_{\rm pt}\tensor\B_{\rm pt}\rightarrow \B,$ where
$\widehat \Psi( G_1\tensor G_2) (\q) := G_1(\q^1)G_2(\q^2),$ $\q^1$ and 
$\q^2$ are defined by $\Psi^{-1}\q= \q^1 e_1 +\q^2 e_2.$
According to \cite{G3}, the ancestor 
potential is given by the following formula:
\beq\label{ancestor}
\A_\tau =\widehat \Psi \widehat R e^{\widehat U/z} 
(\D_{\rm pt}\tensor\D_{\rm pt}),
\eeq  
where $R$ and $e^{U/z}$ are certain respectively upper-triangular and 
lower-triangular linear transformations in $\L^{(2)}\lieGL(\C^2).$ 

The factor $e^{U/z}$ 
in formula \eqref{ancestor} is redundant because $\D_{\rm pt}$ satisfies the 
string equation (\cite{G3}). It will be explained in subsection 3.1 that 
$\D_{\rm pt}=\A_{\rm pt}.$ 
Also, $\widehat \Psi$ intertwines the action of the twisted loop
groups $\L^{(2)}\lieGL(\C^2)$ and $\L^{(2)}\lieGL(H)$ i.e.,
$\widehat \Psi \widehat R \widehat \Psi^{-1} = (\Psi R\Psi^{-1})\sphat.$
Thus formula \eqref{ancestor} is equivalent to:
\beq\label{ancestor_potential}
\A_\tau(\q) = 
\widehat R 
\(\A_{\rm pt} (\q^1)\A_{\rm pt} (\q^2)\),
\eeq
where $R\in \L^{(2)}\lieGL(H)$ is a certain upper-triangular linear 
transformation and
$\q^i$ are the coordinates of $\q$ with respect to the basis
$\{{\bf 1}_1,{\bf 1}_2\}.$

The linear transformation $R$ can be expressed in terms of oscillating
integrals. 
Let $\gb_i(\infty)\subset \C^*$ be an extension of the Lefschetz thimble 
$\gb_i$ along a path $C(\infty)$ starting at $\gl_0$ and approaching 
$\gl=\infty$ in such a way that $\Re \gl < 0.$ Then the oscillating
integrals
\beq\label{osc_i}
\J_i(t,z) =(-2\pi z)^{-1/2} \int_{\gb_i(\infty)} e^{f_t(x)/z}\omega,\quad
i=1,2,
\eeq
are well defined and according to \cite{G3} the matrix 
\ben
J=
\begin{bmatrix}
\J_1      & \J_2 \\
z\d_t\J_1 & z\d_t \J_2
\end{bmatrix} 
\cdot
\begin{bmatrix}
e^{-u_1/z}      & 0 \\
 0 & e^{-u_2/z}
\end{bmatrix}
\een
is asymptotic as $z\rightarrow 0$ to the matrix of the 
linear operator $R$ with respect to the basises $\{{\bf 1}_1,{\bf 1}_2\}$
and $\{\phi^0,\phi^1\}$ respectively in the domain and the co-domain of $R.$

\subsection{Conjugating vertex operators}
Let $\gl$ be sufficiently close to the critical value $u_i$ and 
assume that the path $C$ specifying the vertex operator 
$\Gamma_\tau^{\gb_i/2}$ is the same as in \prref{I_ui}. 
According to \eqref{conj_R}, in order 
to conjugate the vertex operator $\Gamma_\tau^{\gb_i/2}$ by the 
symplectic transformation $R$, we need to derive formulas 
for the vector $R^{-1}\f_\tau^{\gb_i/2}$ and for the  
phase factor $V(\f_\tau^{\gb_i/2})_-^2/2.$ The computations 
are essentially the same as in \cite{G1}. 

\begin{lemma}\label{laplace}
a) The period vectors satisfy:
$\d_\gl I^{(n)}_{\gb_i} = I^{(n+1)}_{\gb_i},$

b) For every $k\geq 0,$ the following formula holds:
\ben
(-z)^{k+3/2}\J_i = \frac{1}{\sqrt{2\pi}}
\int_{u_i}^{\infty} e^{\gl/z}
\(\int_{\gb_i(\gl)} \frac{(\gl-f_t)^k}{k!} \omega\) d\gl ,
\een
where the integration path is $C(\infty)\circ C_i^{-1}.$
\end{lemma}
\proof
a)
Note that
\ben
\int_{\gb_i(\gl)}\frac{(\gl-f_\tau(x))^{k+1}}{(k+1)!}\omega =
\int_{\gl_0}^{\gl} \(\int_{\d \gb_i(\xi)}
\frac{(\gl-f_\tau(x))^{k+1}}{(k+1)!}\frac{\omega}{df_t}\) d\xi. 
\een
Differentiating with respect to $\gl$ we get
\ben
&&
\d_\gl \int_{\gb_i(\gl)}\frac{(\gl-f_\tau(x))^{k+1}}{(k+1)!}\omega =
\int_{\gl_0}^{\gl} \(\int_{\d \gb_i(\xi)}
\frac{(\gl-f_\tau(x))^{k}}{k!}\frac{\omega}{df_t}\) d\xi + \\
&&
+\int_{\d \gb_i(\gl)}
\frac{(\gl-f_\tau(x))^{k+1}}{(k+1)!}\frac{\omega}{df_t}= 
\int_{\gb_i(\gl)}\frac{(\gl-f_\tau(x))^{k}}{k!}\omega, 
\een
where we used that the function $\gl-f_t(x)$ vanishes on the 
boundary of the cycle $\gb_i(\gl)\in H_1(\C^*,f_t^{-1}(\gl);\Z).$
Part a) follows.

b) 
By definition, 
\ben
(-2\pi z)^{1/2}\J_i & =& 
\int_{u_i}^{-\infty} e^{\gl/z} 
\int_{\d \gb_i(\gl)} \frac{\omega}{df_t} d\gl = 
\int_{u_i}^{-\infty} e^{\gl/z} 
\d_\gl\left( \int_{\d \gb_i(\gl)} d^{-1}\omega\right) d\gl =  \\
& &
\int_{u_i}^{-\infty} e^{\gl/z} 
\d_\gl\left(\int_{ \gb_i(\gl)} \omega\right) d\gl = 
-z^{-1}\int_{u_i}^{-\infty} e^{\gl/z} 
\left(\int_{ \gb_i(\gl)} \omega \right)d\gl.
\een
Part b) follows from a) and integration by parts.
\qed

\begin{lemma}\label{vector}
The following formula holds:
\beq\label{psi_r}
R^{-1} \f_\tau^{\gb_i/2} = 
\sum_{n\in \Z} (-z\d_\gl)^n \frac{{\bf 1}_i}{\sqrt{2(\gl-u_i)}}, 
\eeq
where (for $n<0$) $\d_\gl^{-1}$ means integration and the corresponding
integration constants are ``set to $0$''.
\end{lemma} 
\proof
Let $J_i = 
e^{-u_i/z}\J_i\phi^0 +e^{-u_i/z}(z\d_t\J_i)\phi^1.$ Using 
\leref{laplace} b) we get 
\ben
J_i = (-2\pi z)^{-1/2} 
\int_{u_i}^{-\infty} e^{(\gl-u_i)/z}I_{\gb_i}^{(0)}(\gl,\tau)d\gl.
\een
From here our argument is the same as the proof of Theorem 3 in 
\cite{G1}.
Near the critical value the period  $I_{\gb_i}^{(0)}(\gl,\tau)$ 
has the expansion \eqref{i_0}. Using the change of variables 
$\gl-u_i = -zx^2/2$ we compute
\ben
\frac{2}{\sqrt{-2\pi z}}
\int_{u_i}^{-\infty} e^{(\gl-u_i)/z}
[2(\gl-u_i)]^{k-1/2} d\gl & = & (-z)^k\frac{1}{\sqrt{2\pi}}
\int_{-\infty}^\infty e^{-x^2/2}x^{2k}dx = \\
&=& (-z)^k (2k-1)!! \ . 
\een  
Thus $J_i$ has the following asymptotic
\ben
J_i \sim \sum_{k=0}^\infty (2k-1)!!\, A_{i,k} (-z)^k. 
\een
Since, by definition, the asymptotic of $J_i$ is $R{\bf 1}_i$ we get 
$A_{i,k}=(-1)^kR_k/(2k-1)!! $. Thus
{\allowdisplaybreaks
\ben
\f_\tau^{\gb_i}(\gl) & = & 
\sum_{n\in\Z}(-z\d_\gl)^n I^{(0)}_{\gb_i}(\gl,\tau) = 
2\sum_{n\in \Z}\sum_{k=0}^\infty 
(-z\d_\gl)^n (-1)^kR_k{\bf 1}_i \frac{[2(\gl-u_i)]}{(2k-1)!!}^{k-1/2} = \\
&=&
\sum_{n\in\Z}(-z\d_\gl)^n  R_k{\bf 1}_i 
(-\d_\gl)^{-k} \frac{1}{\sqrt{2(\gl-u_i)}} = 
2 R \sum_{n\in\Z}(-z\d_\gl)^n\frac{{\bf 1}_i }{\sqrt{2(\gl-u_i)}}.
\een}
\qed

\begin{lemma}\label{phase_factor}
Let $V$ be the quadratic form \eqref{conj_v}. Then
\beq\label{phase}
V\f_-^2 = 
-\lim_{\ge\rightarrow 0}
\int_{\gl}^{u_i+\ge}
\(
\( I_\ga^{(0)}(\xi,\tau) , I_\ga^{(0)}(\xi,\tau)\)
-\frac{1}{ 2(\xi-u_i) } 
\) d\xi , 
\eeq
where $\f:=\f_\tau^{\gb_i/2}$ and the integration path is 
$C_i(\ge)\circ C^{-1}.$ 
\end{lemma}
\proof 
The proof is taken from \cite{G1}, page 490. 
When 
$\f=\sum_{k\in\Z} I_{\gb_i/2}^{(k)}(-z)^k$ we have 
$\f_-=\sum_{k\geq 0}(I_{\gb_i/2}^{(-1-k)},\phi^a)q_{k,a}.$
Using $\d_\gl I_{\gb_i}^{(-1-k)}=I_{\gb_i}^{(-k)},$ we find  
\ben
\d_\gl V\f_-^2 
&=& \frac{1}{4}\sum_{k,l\geq 0} \d_\gl 
\(V_{k,l}I_{\gb_i}^{(-1-l)},I_{\gb_i}^{(-1-k)}\) = 
\frac{1}{4}\sum_{k,l\geq 0} 
\( [V_{k-1,l} + V_{k,l-1} ] I_{\gb_i}^{(-l)},I_{\gb_i}^{(-k)} \)=\\
&=&
\frac{1}{4}(I_{\gb_i}^{(0)},I_{\gb_i}^{(0)})- 
\frac{1}{4}
(\sum_{l\geq 0}R_l^*I_{\gb_i}^{(-l)},\sum_{k\geq 0}R_l^*I_{\gb_i}^{(-k)})  .
\een
On the other hand, $R^*(z) = R^{-1}(-z)$ because $R\in \L^{(2)}\lieGL(H).$
Thanks to \leref{vector} 
$\sum_{k\geq 0}R_l^*I_{\gb_i}^{(-k)}=
2{\bf 1}_i/\sqrt{2(\gl-u_i)}.$ Also $V\f_-^2 = 0$ at $\gl=u_i$ because 
$I^{(-1-k)}_{\gb_i} = 2[2(\gl-u_i)]^{k+1/2}({\bf 1}_i +\ldots )$ vanish
at $\gl=u_i.$ The lemma follows. 
\qed

\subsection{Proof of \thref{t1}}
According to the discussion in subsection 2.2, it is enough to show 
that the HQE \eqref{HQE_a} has no pole at $\gl=u_i,\ i=1,2.$
Let us assume that $\gl$ is close to the critical value $u_i$ and that
the path $C$ between $\gl_0$ and $\gl$ 
is the same as in \leref{I_ui}. There are two cases:

{\em Case 1.} When $i=1$ i.e., $\gl$ is close to $u_1.$
Using formula \eqref{ancestor_potential} for the ancestor potential,  
the conjugation formula
\eqref{conj_R}, \leref{vector}, and \leref{phase_factor}, we transform
the HQE \eqref{HQE_a} into 
\beqa \label{conj_HQE}
&&
\(\Gamma_\tau^{\gd\#}\tensor \Gamma_\tau^\gd \)
\(\widehat R\tensor \widehat R\) \\
&& \notag
\[ b_\ga \( 
\Gamma_{u_1}^{+ }\tensor \Gamma_{u_1}^{-} -
  \Gamma_{u_1}^{-}\tensor \Gamma_{u_1}^{+}\)  \right. 
\left. 
\(\A_{\rm pt}(\q^1)\tensor \A_{\rm pt}(\q^1)\)\  {d\gl}\] 
\(\A_{\rm pt}(\q^2)\tensor \A_{\rm pt}(\q^2)\),
\eeqa
where $\Gamma_{u_1}^\pm$ are the vertex operators \eqref{vop_kdv} 
and the function $b_\ga$ is given by
\ben
b_\ga = c_\ga e^{V\f_-^2}=\lim_{\ge\rightarrow 0} \exp\(
-\int_1^\ge \frac{d\xi}{2\xi} + 
\int_\gl^{u_1+\ge}\frac{d\xi}{2(\xi-u_1)} \) =  
\exp\(\int_\gl^{u_1+1}\frac{d\xi}{2(\xi-u_1)} \).
\een
Note that the expression in the $[\ \ ]$-brackets in \eqref{conj_HQE} 
is precisely the HQE \eqref{KdV} of the ancestor potential of a point. 
Thus it is regular in $\gl.$  On the other hand 
$I_\phi^{(n)}(\gl,\tau)$ is polynomial in 
$\gl$ if $n<0$ and it is $0$ otherwise (see \eqref{I_phi} and 
\eqref{I_phii}). Thus $\Gamma_\tau^\delta$ is regular in $\gl.$

{\em Case 2.} When $i=2$ i.e., $\gl$ is close to $u_2.$ We reduce 
this case to the previous one. Let $\ga'=\gb_2/2$ and
\ben
c_{\ga'}= \lim_{\ge\rightarrow 0}
\exp\( \int^{u_2+\ge}_\gl 
\(I^{(0)}_{\ga'}(\xi,\tau),I^{(0)}_{\ga'}(\xi,\tau)\)d\xi
-\langle \ga',\ga' \rangle\,\int_1^\ge\frac{d\xi}{\xi}  \).
\een
We will prove that $c_\ga(\gl)/c_{\ga'}(\gl)$ is a constant 
depending only on the reference point $\gl_0$ and the paths 
$C_1$ and $C_2.$

For every $\chi = r_1\gb_1 + r_2\gb_2,\ r_1,r_2\in \QQ$ let 
$\W_\chi= (I_\chi^{(0)}(\xi,\tau),I_\chi^{(0)}(\xi,\tau))d\xi$ 
 -- it is a meromorphic 1-form with simple poles at $\xi=u_1$ and 
$\xi=u_2$ (see \leref{I_ui}). 
Let $C_i(\ge)$ be the path starting at $\gl_0$  and reaching the 
point $u_i+\ge$ along $C_i.$ Then
\beq\label{const}
c_\ga(\gl)/c_{\ga'}(\gl) = \lim_{\ge\rightarrow 0} 
\exp\( \int_{C_1(\ge)\circ C} \W_\ga - \int_{C_2(\ge)\circ C}\W_{\ga'}\)=
\lim_{\ge\rightarrow 0} 
\exp\(
\int_{ C_1(\ge)\circ C_2^{-1}(\ge) } \W_\ga \),
\eeq
where for the second equality we used that $\W_\ga = \W_{\ga'}$ which
follows from $\ga=\phi/2 - \ga'$ and $I_\phi^{(0)} = 0.$ 

On the other hand
$
\Gamma_\tau^{\pm\ga} = \Gamma^{\pm\phi/2}_\tau \Gamma_\tau^{-\ga'}.
$
According to \leref{OPE1}, 
\ben
\(\Gamma_\tau^{\delta \#}\tensor\Gamma_\tau^\delta\)
\(\Gamma_\tau^{\pm\phi/2}\tensor  \Gamma_\tau^{\mp\phi/2}\) =
e^{\pm\pi i(\hat w_\tau' - \hat w_\tau'')}
\Gamma_\tau^{\delta \#}\tensor\Gamma_\tau^\delta =
e^{\pi i(\hat w_\tau' - \hat w_\tau'')}
\Gamma_\tau^{\delta \#}\tensor\Gamma_\tau^\delta ,
\een
where the second equality holds whenever 
$\widehat w_\tau'-\widehat w_\tau''\in \Z.$  
Thus the HQE \eqref{HQE_a} are equivalent to
the similar HQE, with $\ga'$ instead of $\ga.$  

The proof that the HQE corresponding to $\ga'$ are regular at $\gl=u_2$
is the same as in the previous case. 
\qed

\sectionnew{From Ancestors to Descendants}
In this section we give a proof of \coref{c}. 

\subsection{Descendants and ancestors}
Let $S_\tau(z)=1+S_1 z^{-1}+S_2z^{-2}+\ldots$  be the operator 
series defined by
\beq
\label{S}
(S_\tau\phi_\ga,\phi_\gb)=(\phi_\ga,\phi_\gb)+
\sum_{k\geq 0} 
\langle \phi_\ga\psi^k,\phi_\gb \rangle _{0,2}(\tau)z^{-1-k}.
\eeq
It is a basic fact in quantum cohomology theory that $S_\tau$ is a 
lower-triangular 
linear transformation from $\L^{(2)}\lieGL(H)$
(see \cite{G3}, section 6 and the references there in). 
According to \cite{CG}, Appendix 2,
\beq\label{desc_anc}
\D_X = e^{\rm F^{(1)}(\tau)}\widehat S_\tau^{-1} \A_\tau,
\eeq 
where ${\rm F^{(1)}} = \left. \F^{(1)}\right|_{t_0=\tau,t_1=t_2=\ldots =0}$
is the genus-1 no-descendants potential. 
The action of $S$ on a function $G$
from the Fock space $\B$ is given by the following formula 
(\cite{G3}, Proposition 5.3):
\beq\label{S-fock}
\widehat S_\tau^{-1}{G}(\q)  =  e^{W(\q,\q)/2\ge^2}
                {G}([S_\tau\q]_+),
\eeq      
where the quadratic form 
$W(\q,\q)=\sum (W_{kl}q_l,q_k)$ is defined by
\beq
\label{conj-w}
W_{kl}w^{-k}z^{-l} = \frac{S_\tau^*(w)S_\tau(z)-1}{w^{-1}+z^{-1}}.
\eeq
When $X={\rm pt},$ $S_\tau = e^{\tau/z}$ and the genus-1 no-descendants 
potential ${\rm F^{(1)}}$ vanishes for dimensional reasons. Also,
according to the string equation $(1/z)\sphat \ \D_{\rm pt} =0.$ Thus formula
\eqref{desc_anc} yields $\A_{{\rm pt},\tau} = \D_{\rm pt}.$

The proof of \coref{c} amounts to conjugating the vertex operators 
in the HQE \eqref{HQE_a} by $S_\tau.$ 
We will use the following formula
(\cite{G1}, formula (17)):
\beq
\label{S-conjugation}
\widehat S_\tau^{-1} e^{\hat f}\hat S_\tau  = 
e^{-W(   (S_\tau^{-1}f)_+  ,  (S_\tau^{-1}f)_+   )/2}
e^{      (S_\tau^{-1}f) \sphat                      },
\eeq 
where $+$ means truncating the terms corresponding to the 
negative powers of $z.$ The key ingredient in the computation is
that $\f_\tau^{\ga}$ and $S_\tau$ satisfy the 
same system of differential equations with respect to the parameter 
$\tau.$  

\subsection{The small quantum differential equation}

Let us assume that the parameter $\tau = t\,P.$ The quantum multiplication
by $P$ is, by definition, a linear operator on $H,$ defined as follows:
\ben
\( P\bullet_\tau\phi_i\, ,\, \phi_j\) := 
\<P,\phi_i,\phi_j\>_{0,3}(\tau)  
\een 
The linear operator $P\bullet_\tau$ is self-adjoint 
with respect to the {\Poincare} pairing --- the correlator is symmetric,
so we can switch $\phi_i$ and $\phi_j.$ According to 
\cite{HZ}, formula (26.15), $P\bullet_\tau$ has matrix  
$\begin{bmatrix}0 & Qe^t \\ 1 & 0\end{bmatrix}$
with respect to the basis $\{1,P\}.$  

The ordinary differential equation 
\beq\label{qde}
z\d_t\Phi(t) = (P\bullet_\tau)\Phi(t),\quad \Phi(t)\in H
\eeq 
is called {\em the small quantum differential equation} of $\C P^1.$
It admits a solution in terms of Gromov--Witten invariants of $\C P^1$ ---
the operator series $S_\tau$ is a fundamental solution to \eqref{qde}
(\cite{HZ}, Proposition 28.0.2).

Part of the mirror symmetry phenomena is that the small quantum differential
equation can be solved in terms of the oscillating integrals \eqref{osc_i}
of the mirror model of $\C P^1.$ 
Let $J_i(t,z),\ i=1,2$ be vector-valued functions with values in $H$ defined
by:
\ben
(J_i,{\bf 1}) = \J_i,\quad (J_i,P) = z\d_t \J_i,
\een
where $\J_i$ are the oscillating integrals \eqref{osc_i}. 
\begin{lemma}\label{de}
Vectors $J_i,\ i=1,2$ are solutions to the small quantum
differential equation \eqref{qde}. They satisfy
the following homogeneity condition:
\ben
(z\d_z+E)J_i(t,z)=\mu J_i(t,z),
\een
where $E=2\d_t$ is {\em the Euler vector field} and $\mu$ is {\em the Hodge 
grading operator} i.e., in the basis $\{{\bf 1},P\},$ 
$\mu = {\rm diag}\{1/2,-1/2\}.$
\end{lemma}  
\proof
For the first part of the lemma see \cite{G3}, section 10 (or just 
differentiate and use integration by parts).
  
Let us prove the homogeneity condition. Using that $J_i = (J_i,{\bf 1})P+(J_i,P){\bf 1}=
\J_i P + (z\d_t\J_i){\bf 1}$ we see that the equation is equivalent to
\ben
(z\d_z+E)\J_i = -(1/2)\J_i \quad \mbox{ and } \quad
(z\d_z+E)z\d_t\J_i = (1/2)z\d_t\J_i.
\een
It is enough to prove the 1-st equation because the second one 
can be derived from the 1-st one. 
\ben
&&
(z\d_z+E)\J_i = (z\d_z+2\d_t)(-2\pi z)^{-1/2}
\int_{\gb_i(\infty)}\exp\(f_t/z\)dx/x = \\
&&
(-2\pi z)^{-1/2}\int_{\gb_i(\infty)}
\(-(1/2) + z^{-1}(-f_t + 2 Qe^t/x) \)\exp\(f_t/z\)dx/x =\\
&&
(-1/2)\J_i + (-2\pi z)^{-1/2}\int_{\gb_i(\infty)} d \exp\(f_t/z\)=
(-1/2)\J_i.
\een
\qed

\begin{lemma}{\label{ode}} The series $\f_\tau^{\gb_i}$ satisfies 
the small quantum differential equation \eqref{qde} and the 
homogeneity condition:
\beq\label{hom}
(z\d_z+\gl\d_{\gl}+E)\f_\tau^{\gb_i} = (\mu - 1/2)\f_\tau^{\gb_i}
\eeq
\end{lemma}
\proof 
We will show that $\f_\tau^{\gb_i}$ is a solution to
\eqref{qde}. The derivation of \eqref{hom} is similar.

The series $\f_\tau^{\gb_i}$ satisfies \eqref{qde} if and only if 
$\d_t I_{\gb_i}^{(n)} = -(P\bullet_\tau)I_{\gb_i}^{(n+1)}.$ 
It is enough to prove that the last equality holds for every 
$n=-k-1,$ $k\geq 0$ --- according to \leref{laplace}, a), 
for $n=k, \ k\geq 0$ we need only to 
differentiate $k+1$ times, the equality corresponding to $n=-1.$
Thus we need to prove that
\beq \label{de_tt}
(\d_t I^{(-1-k)}_{\gb_i},{\bf 1}) = -(P\bullet_\tau I^{(-k)}_{\gb_i},{\bf 1})
\quad \mbox{ and }\quad
(\d_t I^{(-1-k)}_{\gb_i},P) = -(P\bullet_\tau I^{(-k)}_{\gb_i},P)
\eeq
On the other hand, by definition, 
$$I^{(-1-k)}_{\gb_i}=(I^{(-1-k)}_{\gb_i},{\bf 1})P + 
(I^{(-1-k)}_{\gb_i},P){\bf 1} = 
\(\d_\gl\mathcal{I}_i^{(k+1)}\) P+ \(-\d_t \mathcal{I}_i^{(k+1)}\){\bf 1},$$ 
where
\ben
\mathcal{I}_i^{(k+1)} = 
\int_{\gb_i(\gl)} \frac{(\gl-f_t)^{k+1}}{(k+1)!} \omega.  
\een
Thus after a short computation, we get that the differential 
equations \eqref{de_tt} are equivalent to a single differential 
equation: 
\beq\label{de_i}
-\d_t^2 \mathcal{I}_i^{(k+1)} = 
(P\bullet_\tau P, {\bf 1})\d_t \d_\gl \mathcal{I}_i^{(k+1)} -
    (P\bullet_\tau P,P)\d_\gl^2 \mathcal{I}_i^{(k+1)}.
\eeq

Similarly, the fact that $J_i$ is a solution to the small quantum
differential equation is equivalent to a single
differential equation for the oscillating integral $\J_i:$
\ben
\d_t^2\J_i = (P\bullet_\tau P,{\bf 1})(1/z)\d_t\J_i + 
(P\bullet_\tau P,P)(1/z^2)\J_i.
\een
According to \leref{laplace} b), $(-z)^{k+5/2}\J_i$ is a Laplace 
transform along the path from $\mathcal{I}_i^{(k+1)}.$ Thus the last equation 
implies \eqref{de_i} because under the Laplace transform multiplication by 
$1/z$ corresponds to $-\d/\d\gl.$
\qed

\subsection{The period vectors in a neighborhood of $\gl=\infty$}
Let us assume that the reference point $\gl_0$ belongs to a 
neighborhood of $\infty,$ i.e. $\gl_0$ is outside a disk containing the 
two critical points $u_1$ and $u_2.$ Let $\gl$ and the path 
$C$ specifying $\gb_i(\gl),$ $i=1,2$ belong to this neighborhood of
$\infty$ as well. We will show that each period vector expands into a series 
of the following type:
\beq\label{periods_expansion}
\Big(\ \sum_{j=0}^N A_j(\tau)\gl^j\ \Big) \log \frac{\gl}{\sqrt Q} + 
\sum_{k\geq 0} B_k(\tau)\gl^{-k}\ ,
\eeq
where the coefficients are vectors in $H$ which depend polynomially on
$t$ and $Qe^t.$

The equation $f_t(x)=\gl$ has two solutions near $\gl=\infty$, which can 
be expanded into Laurent series in $\gl^{-1}$ as follows:
\ben
&&
x_+  = \gl+ a_0 + a_1\gl^{-1} + \ldots \\
&&
x_-  =   Qe^t\gl^{-1} + 
        b_2\gl^{-2} +b_3\gl^{-3} +\ldots , 
\een 
where the coefficients are polynomials in $t$ and $Qe^{t}.$ 

The boundary of the cycle 
$\gb_i(\gl)\in H_1(\C^*,f_t^{-1}(\gl);\QQ)$ is given by:
\beq
\label{d_alpha}
\d \gb_i(\gl) = \pm \frac{1}{2}([x_+]-[x_-]),
\eeq
where the sign depends on the choice of the path $C_i$ 
(the sign does not depend on $C$!). Let us assume that the sign is 
$+$ for $i=1$ and $-$ for $i=2,$ otherwise the argument is similar.
Using the definition we find the following formula: 
\beq
\label{period_-1}
I_{\gb_i}^{(- 1)}(\gl,\tau)  =   
\pm\frac{1}{2}\log \frac{x_+}{x_-}\, P\ \pm \frac{1}{2}(x_+-x_-)\, {\bf 1}   .
\eeq
Substituting the Laurent expansions of $x_+$ and $x_-$ in \eqref{period_-1}
we see that, near $\gl=\infty$, the period $I^{(-1)}_\ga$ expands into a 
series of the type \eqref{periods_expansion}.

The period vector $I_{\gb_i}^{(n)},\ n>0$ has an 
expansion of type \eqref{periods_expansion} obtained by differentiating 
$n+1$ times (with respect to $\gl$)  
the expansion of $I_{\gb_i}^{(-1)}.$ When $n<-1$,
due to \leref{ode}, the following recursive relation holds:
\beq\label{periods_recursion}
(\gl-2P\bullet)\,   I_{\gb_i}^{(-k  )}(\gl,\tau) = 
(\mu+k+\frac{1}{2})I_{\gb_i}^{(-k-1)}(\gl,\tau).
\eeq
When $k\geq 1$ the linear operator $\mu+k+(1/2)$ is invertible. Thus 
$I_{\gb_i}^{(-k-1)}$ can be expressed in terms of $I_{\gb_i}^{(-k)}.$
Arguing by induction we find that $I^{(n)}_{\gb_i},\ n<0$ expands into 
a series of type \eqref{periods_expansion}.

Denote 
$$\f^{\gb_i}(\gl)=
\sum_n I_{\gb_i}^{(n)}(\gl)(-z)^n:= S_\tau^{-1}\f_\tau^{\gb_i}(\gl).$$
Each coefficient  $I_{\gb_i}^{(n)}(\gl)$ is a formal series 
of type \eqref{periods_expansion} with coefficients independent of $\tau.$ 
The last statement follows from the fact that 
$\f_\tau^{\gb_i}$ and  $S_\tau$ satisfy the same differential 
equation with respect to $t.$ Furthermore, using that 
$S_\tau^{-1}(z)=S_\tau^*(-z)$, we get
\ben
I_{\gb_i}^{(n)}(\gl) =I_{\gb_i}^{(n)}(\gl,\tau) + 
S_1^* I_{\gb_i}^{(n+1)}(\gl,\tau) + \ldots =
\(1+S_1^*\d_\gl +S_2^*\d_\gl^2 + \ldots \)I_{\gb_i}^{(n)}(\gl,\tau). 
\een
The last expression is a series of type \eqref{periods_expansion}
whenever $I_{\gb_i}^{(n)}(\gl,\tau)$ is such. 

For every $\ga=r_1\gb_1+r_2\gb_2,$ $r_1,r_2\in\QQ$ define 
\ben
I_\ga^{(n)}(\gl) = r_1I_{\gb_1}^{(n)}(\gl) +r_2I_{\gb_2}^{(n)}(\gl),\quad
\f^\ga(\gl) =\sum_n I_\ga^{(n)}(\gl)(-z)^n,\quad
\Gamma^{\ga}(\gl) = e^{\hat \f^{\ga}(\gl)}.
\een
\begin{lemma}\label{vector_S} Let $\ga=\gb_1/2$. Then the following 
formulas hold:
\ben
&&
I_\ga^{(-1-n)}(\gl) = 
\frac{\gl^n}{n!} (\log \frac{\gl}{\sqrt Q} -\CC_n)P+
\frac{\gl^{n+1}}{2(n+1)!} ,\ n\geq 0\ , \\
&&
I_\ga^{(0)}(\gl)=\frac{1}{\gl} P +\frac{1}{2},\ \ \ \  
I_\ga^{(n)}(\gl)=\frac{(-1)^{n}n!}{\gl^{n+1}} P ,\ n>0 \ .
\een
\end{lemma}
\proof
The coefficients $S_k,k\geq 1$ in front of the powers of $z$ in the 
series $S_\tau$  depend polynomially on $t$ and  $Qe^t$ 
and after letting $t=Q e^t =0$  they all vanish 
(see \cite{HZ}, Exercise 28.1.1.). Thus:
\ben
\f^\ga(\gl) = \f^\ga(\gl,\tau) |_{t=Qe^t=0}. 
\een  
In particular  we find: 
\beq\label{p-limit}
I_\ga^{(-1)}(\gl) = 
\log \frac{\gl}{\sqrt Q}\, P +\frac{\gl}{2}\, {\bf 1 }  .
\eeq
The other vectors $I_\ga^{(n)}$ can be easily obtained from
$I_\ga^{(-1)}.$ Indeed, for $n=k\geq 0$  
differentiate  $k+1$ times \eqref{p-limit} and for $n=-k-1,k\geq 1$  
use the recursive relation \eqref{periods_recursion} specialized to 
$t=Qe^t=0:$
\ben
(\gl-2P\cup)I_{\ga}^{(-k)} = (\mu+k+1/2)I_{\ga}^{(-k-1)},
\een 
where $P\cup$ is the classical cup product multiplication by $P$ in 
the cohomology algebra $H.$  
\qed

The phase factor $W_\tau(\f_+^\ga(\gl),\f_+^\ga(\gl))$ can be computed
as follows: 
{\allowdisplaybreaks
\ben
&&
\frac{d}{d\xi} W_\tau(\f_+^\ga(\xi),\f_+^\ga(\xi)) = 
-\sum_{k,l\geq 0} 
\( (W_{k,l-1}+W_{k-1,l})(-1)^l I_\ga^{(l)}(\xi), 
                        (-1)^k I_\ga^{(k)}(\xi) ) \) = \\
&&
-\sum_{k,l\geq 0} 
\( S_l(-1)^l I_\ga^{(l)}(\xi), 
                        S_k(-1)^k I_\ga^{(k)}(\xi) ) \) 
+\( I_\ga^{(0)}(\xi), I_\ga^{(0)}(\xi) \) = \\
&&
-\(I_\ga^{(0)}(\xi,\tau),I_\ga^{(0)}(\xi,\tau)\) +
\( I_\ga^{(0)}(\xi), I_\ga^{(0)}(\xi)\) 
\een}
where for the first equality we used \leref{laplace} a) and for the second 
one -- the definition \eqref{conj-w} of the quadratic form $W_\tau$.  
Since $\f^\ga_+=(1/2){\bf 1}$ at $\xi=\infty$ we get 
\ben
W_\tau(\f_+^\ga(\gl),\f_+^\ga(\gl)) = 
\frac{1}{4}W_\tau({\bf 1},{\bf 1}) +
\int^\infty_\gl \( 
(I_\ga^{(0)}(\xi,\tau),I_\ga^{(0)}(\xi,\tau)) - 
( I_\ga^{(0)}(\xi), I_\ga^{(0)}(\xi)) \)d\xi,
\een 
where the integration path is $C(\infty)\circ C^{-1}.$

The term $W_\tau({\bf 1},{\bf 1})$ can be computed as follows:
\ben
&&
W_\tau({\bf 1},{\bf 1})=(W_{0,0}{\bf 1},{\bf 1}) = (S_1{\bf 1},{\bf 1}) = 
\sum_{d,n} \frac{Q^d}{n!}
\langle {\bf 1}, {\bf 1}, \tau,\ldots,\tau \rangle_{0,n+2,d} =  \\
&&
\langle {\bf 1},{\bf 1},\tau \rangle_{0,3,0} = \int_{[\C P^1]}\, \tau = t,
\een
where thanks to the string equation all terms in the sum on the first line, 
except the ones corresponding to $d=0$ and $n=1$, vanish.  

\begin{lemma}\label{conj_fa}The following formula holds:
\ben
c_\ga(\gl)
\(\Gamma_\tau^{\pm\ga}\tensor \Gamma_\tau^{\mp\ga} \)
\(\widehat S_\tau \tensor \widehat S_\tau \) = 
\frac{b_\ga(t)}{\gl}
\(\widehat S_\tau \tensor \widehat S_\tau \)
\(\Gamma^{\pm\ga}\tensor \Gamma^{\mp\ga} \),
\een
where the function $b_\ga(t)$ depends only on the choice of the 
reference point $\gl_0$ and the paths $C_i,$ $i=1,2$ and 
$C(\infty).$ 
\end{lemma}
\proof
According to the conjugation formula \eqref{S-conjugation},
\ben
c_\ga(\gl)
\(\Gamma_\tau^{\pm\ga}\tensor \Gamma_\tau^{\mp\ga} \)
\(\widehat S_\tau \tensor \widehat S_\tau \) = 
B_\ga
\(\widehat S_\tau \tensor \widehat S_\tau \)
\(\Gamma^{\pm\ga}\tensor \Gamma^{\mp\ga} \),
\een
where the function $B_\ga$ is given by the following formula:
{\allowdisplaybreaks
\ben
&
B_\ga  = &\lim_{\ge\rightarrow 0} \exp \left (
\int_{C^{-1}\circ C_1(\ge)} 
\(I_\ga^{(0)}(\xi,\tau),I_\ga^{(0)}(\xi,\tau)\) d\xi
-(1/2) \int_1^\ge \frac{d\xi}{\xi} -  \right. \\
&&
\left. 
-\frac{t}{4}-
\int_{C(\infty)\circ C^{-1} }  \( 
(I_\ga^{(0)}(\xi,\tau),I_\ga^{(0)}(\xi,\tau)) - 
( I_\ga^{(0)}(\xi), I_\ga^{(0)}(\xi)) \)d\xi
\right) = \\
&
=& e^{-t/4}
\lim_{\ge\rightarrow 0} \exp \left (
\int_{C_1(\ge)} 
\(I_\ga^{(0)}(\xi,\tau),I_\ga^{(0)}(\xi,\tau)\) d\xi
-(1/2) \int_1^\ge \frac{d\xi}{\xi}   \right)\times \\
&&
\times \exp\left( 
-\int_{\gl_0 }^\infty  \( 
( I_\ga^{(0)}(\xi,\tau),I_\ga^{(0)}(\xi,\tau)) - 
( I_\ga^{(0)}(\xi     ),I_\ga^{(0)}(\xi     )) 
\)d\xi  \right)\times\\
&&
\times\exp\left(
- \int_{\gl_0}^\gl ( I_\ga^{(0)}(\xi), I_\ga^{(0)}(\xi)) d\xi \right)
\een}
Let us discuss each of the terms in the last equality. 
The first exponential factor is precisely $c_\ga(\gl_0).$ 

The second exponential factor is  some function on $t,$ which depends on the 
path $C(\infty)$ connecting $\gl_0$ and $\infty.$ The corresponding 
integral is convergent for the following reason:
\ben
I_\ga^{(0)}(\xi,\tau) = \(1-S_1\d_\xi+S_2\d_\xi^2\pm\ldots \)
I_\ga^{(0)}(\xi).
\een 
According to \leref{vector_S}, $I_\ga^{(0)}(\xi)=(1/\xi)P +(1/2)1$. Hence
$$
( I_\ga^{(0)}(\xi,\tau),I_\ga^{(0)}(\xi,\tau)) - 
( I_\ga^{(0)}(\xi     ),I_\ga^{(0)}(\xi     )) = O(\xi^{-2}).
$$
The last exponential factor is equal to $\gl_0/\gl.$ The function 
$b_\ga$ is given by the following formula
\ben
b_\ga(t) = e^{-t/4}\lim_{R\rightarrow\infty}\lim_{\ge\rightarrow 0} 
\exp\left( 
-\int_{u_1+\ge }^R  \( 
( I_\ga^{(0)}(\xi,\tau),I_\ga^{(0)}(\xi,\tau)) \)d\xi
+\int_1^R\frac{d\xi}{\xi} + \int_1^\ge \frac{d\xi}{2\xi}   \right) 
\een
\qed

Let us conjugate the vertex operators $\Gamma_\tau^\delta.$ 
If we change the path $C$ by precomposing it with a loop
around $\gl=\infty,$ then the corresponding thimble $\ga(\gl)$
is transformed into $\ga(\gl) + \phi(\gl).$ Thus using 
\leref{vector_S}, we get the following formula:
\beq\label{vector_f}
\f^{\phi}(\gl) = 2\pi i\ \sum_{n\geq 0} \frac{\gl^n}{n!}P(-z)^{-n-1}.
\eeq
Introduce the vertex operator
\beq\label{vop_d}
\Gamma^\delta = 
\exp\left\{\(\f^{\phi/2\pi i} 
           -P(-z)^{-1}\)\sphat\  (\ge\d_x-\log\sqrt Q)
    \right\}
\exp\left\{(x/\ge)\hat {\bf 1} \right\} .
\eeq
\begin{lemma}\label{conj_fd}
The following formula holds:
\ben
\(\Gamma_\tau^{\delta\#}\tensor \Gamma_\tau^\delta\)
\(\widehat S_\tau\tensor \widehat S_\tau \) = 
e^{-t x^2/(2\ge^2)} 
\(\widehat S_\tau\tensor \widehat S_\tau \)
\(\Gamma^{\delta\#}\tensor \Gamma^\delta\)
e^{-t x^2/(2\ge^2)}. 
\een
\end{lemma}
\proof
Note that
$w_\tau=S_\tau \, P(-z)^{-1}$ and $v_\tau = S_\tau {\bf 1}.$
According to \eqref{S-conjugation},
\ben
&&
\widehat S_\tau^{-1} \Gamma_\tau^\gd \widehat S_\tau = 
\widehat S_\tau^{-1} \exp\left\{
\(\f_\tau^{\phi/2\pi i} -w_\tau\)\sphat\ (\ge\d_x-\log\sqrt Q)
\right\}
\exp\left\{(x/\ge)\hat v_\tau\right\}
\hat S_\tau = \\
&&
\exp\left\{\(\f^{\phi/2\pi i} 
           -P(-z)^{-1}\)\sphat\ (\ge\d_x-\log\sqrt Q)
    \right\}
\exp\left\{(x/\ge)\hat {\bf 1} \right\} 
e^{-W_\tau({\bf 1},{\bf 1})\frac{x^2}{2\ge^2}}  
\een
i.e., $\widehat S_\tau^{-1} \Gamma_\tau^\gd \widehat S_\tau=
\Gamma^\delta e^{-tx^2/(2\ge^2)}.$ Similarly, 
$\widehat S_\tau^{-1} \Gamma_\tau^{\gd\#} \widehat S_\tau=
e^{-tx^2/(2\ge^2)}\Gamma^{\delta\#}.$ 
\qed

\subsection{Proof of \coref{c}}
Let $\omega_\A$ and $\omega_{\D}$ be the 1-forms respectively 
\eqref{HQE_a} and  
\ben
b_\ga \(\Gamma^{\delta\#}\tensor \Gamma^\delta\)
\(\Gamma^\ga\tensor \Gamma^{-\ga} - \Gamma^{-\ga}\tensor \Gamma^{\ga} \)
\(\D(\q') \D(\q'')\)
\frac{d\gl}{\gl}
\een  
According to \leref{conj_fa} and \leref{conj_fd}, and formula \eqref{desc_anc}
for the descendants in terms of the ancestors, 
\ben
e^{   2{\rm F}^{(1)}   (\tau)}\omega_\A=
e^{-tx^2/(2\ge^2)}
\( \widehat S_\tau \tensor \widehat S_\tau  \) \omega_\D
e^{-tx^2/(2\ge^2)}.
\een   
Using formula \eqref{S-fock} for the action of 
$\widehat S_\tau$ on the Fock space, we find that 
up to factors independent of $\gl$ the 1-forms 
$\omega_\D(\q',\q'';\gl)$ and  
$\omega_\A([S_\tau\q']_+,[S_\tau\q'']_+;\gl)$ coincide. 
On the other hand, since $S_\tau$ is a symplectic 
transformation of $\H$ and $w_\tau = S_\tau (P(-z)^{-1}),$ we get
\ben
\hat w_\tau([S_\tau\q']_+) - \hat w_\tau([S_\tau\q'']_+) &=& 
\ge^{-1} \Omega \([S_\tau\q']_+-[S_\tau\q'']_+,w_\tau\) = \\
=\ge^{-1}\Omega \(S_\tau\q'-S_\tau\q'',w_\tau\) &=& 
\ge^{-1}\Omega \(\q'-\q'', P(-z)^{-1} \) = (q_{0,0}''-q_{0,0}')/\ge.
\een
Thus according to \thref{t1}, the total descendant potential satisfies the 
following HQE: for each integer $m$ the 1-form
\ben
\(\Gamma^{\delta\#}\tensor \Gamma^\delta\)
\(\Gamma^\ga\tensor \Gamma^{-\ga} - \Gamma^{-\ga}\tensor \Gamma^{\ga} \)
\(\D(\q') \D(\q'')\)\frac{d\gl}{\gl},
\een
when computed at $q'_{0,0}-q''_{0,0}=m\ge$ is regular in $\gl.$ 
These are precisely the HQE of the Extended Toda Hierarchy which 
were introduced in \cite{M}. According to Theorem 1.1 in \cite{M},
the total descendant potential of $\C P^1$ is a tau-function 
of the Extended Toda Hierarchy. 
\qed
 

\end{document}